\pdfoutput=1
\documentclass[submission,copyright,creativecommons]{eptcs}
\providecommand{\event}{ICLP 2021} 
\usepackage{underscore}           

\usepackage{graphicx}
\usepackage{mathptmx}
\usepackage{tikz-qtree}
\usepackage{filecontents}
\usepackage{csvsimple}
\usepackage{amsfonts,amsmath}
\usepackage{url}
\usepackage{verbatim}
\usepackage{color}

\definecolor{lgray}{gray}{0.95}
\definecolor{lblue}{rgb}{0.90,0.90,1.00}
\definecolor{lyellow}{rgb}{1.00,1.00,0.70}

\usepackage{listings}
\newtheorem{ex}{Example}

\newenvironment{codex}{\small\verbatim}{\endverbatim\normalsize}

\lstnewenvironment{code}
    {\lstset{}%
      \csname lst@SetFirstLabel\endcsname}
    {\csname lst@SaveFirstLabel\endcsname}
\newcommand{\BI}[0]{\begin{itemize}}
\newcommand{\EI}[0]{\end{itemize}}
\newcommand{\I}[0]{\item}
\newcommand{\BE}[0]{\begin{enumerate}}
\newcommand{\EE}[0]{\end{enumerate}}

\newcommand{\BX}[0]{\begin{ex}}
\newcommand{\EX}[0]{\end{ex}}
\newcommand{\BV}[0]{\begin{verbatim}}
\newcommand{\EV}[0]{\end{verbatim}}

\newcommand{\BF}[0]{\begin{filecontents*}{data.csv}}

\newcommand{\BQ}[0]{\color{blue}\begin{quote}}
\newcommand{\EQ}[0]{\end{quote}\color{black}}

\def \bscale1 {0.25}
\def \bscale {0.25}

\usepackage{amssymb}
\usepackage{hyperref}
\usepackage{booktabs}

\begin{document}

\title{Natlog: a Lightweight Logic Programming Language with a Neuro-symbolic Touch}

\author{Paul Tarau
\institute{University of North Texas\\ 
Texas, USA
}
\email{paul.tarau@unt.edu}
}

\def \authorrunning{Paul Tarau}
\def\titlerunning{Natlog: a Lightweight Logic Programming Language with a Neuro-symbolic Touch}

\date{}
\maketitle

\begin{abstract}
We introduce Natlog, a lightweight Logic Programming language, sharing Prolog's unification-driven execution model, but with a simplified syntax and semantics. Our proof-of-concept Natlog implementation is tightly embedded in the Python-based deep-learning ecosystem with focus on content-driven indexing of ground term datasets. As an overriding of our symbolic indexing algorithm, the same function can be delegated to a neural network, serving  ground facts to Natlog's resolution engine. 
Our open-source implementation is available as a Python package at \url{https://pypi.org/project/natlog/}.

{\bf Keyphrases}: {\em 
  Python-based logic programming system,
  embedded logic programming language,
  ground term fact database indexing,
  neuro-symbolic logic programming, 
  logic programming language implementation.
  }
\end{abstract}

\section{Introduction}\label{intro}

With renewed interest in neuro-symbolic AI \cite{d2020neurosymbolic,lamb2020graph,de2020statistical,van2021modular}, integration of logic programming languages into deep-learning ecosystems is becoming of paramount importance. Today's deep-learning frameworks like tensorflow\footnote{\url{https://www.tensorflow.org/}} and torch\footnote{\url{https://pytorch.org/}}  and machine-learning frameworks like scikit-learn\footnote{\url{https://scikit-learn.org/}} \cite{scikit-learn} are all as available as Python packages. 
The need to interoperate with them suggests embedding in the same ecosystem an easy to use, syntactically and semantically lightweight logic programming language, that allows data-scientists  with limited exposure to logic programming to build neuro-symbolic systems with enhanced reasoning abilities. 
While convenient bridges exist between Prolog systems and Python, they require familiarity with the intricacies of an API that assumes user awareness of Prolog's internal term representations and their Python equivalents. Moreover, having a logic engine exposed as a Python callable in the  inner loop of a classifier or regression learner might involve not just robustness issues under multi-threading and multi-processing\footnote{as it is the case with the otherwise excellent pysweep, see \url{https://github.com/damazter/pysweep}}, but also performance, scalability and system deployment issues, all important for practical applications.

The often very large datasets neural networks require for training and inference correspond to ground term databases, sometimes in a flat Datalog format as collected from tabular data or as deeper ground terms coming from JSON, Numpy or Pandas data-frames. As Prolog conflates indexing of non-ground predicates used as {\em code} with indexing of {\em ground fact databases}, it misses opportunities for deep, content-driven indexing  of the latter,  when needed to retrieve relevant facts as efficiently as possible from arbitrarily large datasets or to query neural networks with large models in inference mode.

To address these shortcomings we have developed a proof-of-concept implementation of {\em Natlog}\footnote{\url{https://github.com/ptarau/pypro}}, a simple and practical Prolog-like language,
 with a light, self-explanatory syntax and basic Horn clause + LD-resolution semantics, meant to help data scientists not familiar with Prolog or ASP systems to work with a logic reasoning mechanism modeled on {\em natural language sentences} and easy to embed in the Python-based deep-learning ecosystem. 

We start by listing its key features here, with details expanded in the following sections.
\BI
\I ability to call not only Python functions but also Python generators as if they were just facts in the database
\I ability to pretend to be a Python generator returning a stream of answers
\I ability to yield answers from any point in the resolution process, not just at  its end
\I focus on a clear separation of  ground term databases and rule processing components
\I ability to plug-in a machine-learning subsystem as if it was just a ground dynamic database
\I reliance on pure Python datatypes resulting in ability to be significantly accelerated with the pypy JIT compiler 
\I content-driven indexing, specialized to ground databases
\I natural interface to typical dataset formats (.csv, .json, etc.)
\I Hilog-equivalent semantics, allowing general terms in function and predicate symbol position
\EI

We also propose a simple but novel and practical neuro-symbolic integration mechanism between an LD-resolution driven logic engine and a neural network (or more generally, any comparable Machine Learning tool). {\em For that, we use the neural network as a content-based indexer of a ground-term database}. Besides designing a plugin-mechanism that can replace the default symbolic indexer, we enable the network to return multiple solutions, among which, on the symbolic side, our Prolog engine will ensure, via unification steps, that the process is sound, despite possibly incorrect answers obtained from the network.

The rest of the paper is organized as follows: 
Section \ref{syn} overviews  syntax elements and the Hilog-like semantics of Natlog.
Section \ref{poc} describes details of our proof-of-concept system and its integration in the Python ecosystem.
Section \ref{db} focuses on the content-driven ground database indexing mechanism.
Section \ref{nn} describes details of a plug-in neural network overriding the content-driven database indexer.
Section \ref{rel} discusses related work and
section  \ref{conc} concludes the paper.

Besides familiarity with Horn Clause logic and general ideas about how Prolog is implemented, we assume the reader is familiar with essential Python\footnote{\url{https://www.python.org/}} language constructs and its basic coroutining mechanisms as supported by the {\tt yield} and {\tt yield from} statements.
 
\section{A Natural Syntax with Hilog Semantics}\label{syn}
Our proof-of-concept Natlog implementation relies on Horn Clause logic (in a syntactically lighter form) and a flexible but simple mechanism to delegate to Python everything else. 
\subsection{A syntax warm-up, by examples}
Natlog terms are represented as immutable nested tuples. A parser and a scanner for a simplified Prolog term syntax are used to turn terms into nested Python tuples.
Surface syntax of facts, as read from strings, is just whitespace separated words (with parenthesized tuples) and sentences ended with ``{\tt .}'' or ``{\tt ?}''. Like in Prolog, capitalized tokens denote variables, unless quoted.
With terms  represented as {\em immutable nested tuples}, we adopt a simplified Horn Clause syntax, that removes the need for parenthesizing at clause level but uses parenthesized tuples for deeper terms.

\BX Computing the transitive-closure of a relation. 
\begin{code}
cat is feline.
tiger is feline.
mouse is rodent.
feline is mammal.
rodent is mammal.
snake is reptile.
mammal is animal.
reptile is animal.

tc A Rel C : A Rel B, tc1 B Rel C.

tc1 B _Rel B.
tc1 B Rel C : tc B Rel C.
\end{code}

To query it, at the Python prompt one can type:

\begin{codex}
>>> n=natlog(file_name="natprogs/tc.nat") # load program
>>> n.query("tc Who is animal ?") # execute query
\end{codex}

It will answer with the transitive closure of the ``{\tt is}'' relation:
\begin{codex}
GOAL PARSED: (('tc', 0, 'is', 'animal'),)
ANSWER: ('tc', 'cat', 'is', 'animal')
ANSWER: ('tc', 'tiger', 'is', 'animal')
...
ANSWER: ('tc', 'reptile', 'is', 'animal')
\end{codex}
The computed answers are also available as a Python {\em generator} 
via the method {\tt solve(...)}, 
and an interactive top-level is available via the method {\tt repl()}.
\EX

To match the usual Prolog semantics, lists can be represented as \verb~()~-terminated iterated 2-tuples, with creation of long constant lists delegated to Python.

\BX\label{perm} A simple program generating all permutations of a list.
\begin{code}
perm () ().
perm (X Xs) Zs : perm Xs Ys, ins X Ys Zs.

ins X Xs (X Xs).
ins X (Y Xs) (Y Ys) : ins X Xs Ys.
\end{code}
\EX

The interpreter can handle function and generator calls to Python using a simple prefix operator syntax. It can also call facts in a ground term database and return an answer from an arbitrary position in the code, as summarized by the following prefix annotations to these functions:
\BI
\I \verb|#f A B .. Z|: call f(A,B,C,..,Z) for its side effects, with no result returned.
\I \verb|`f A B .. Z R|: call Python function f(A,B,C,..,Z) and unify R with its result
\I \verb|``f A B .. Z R|: call Python generator f(A,B,C,..,Z) and unify R with its multiple yields, one at a time
\I \verb|^f A B .. Z|: yield term \verb~(f A B ..)~ as an answer from any point in the resolution process
\I \verb|~P A B .. Z|: unify \verb~(P A B .. Z)~with matching facts in the ground fact database

\EI

\subsection{The Terms-as-Nested-Tuples Transformation}

Replacing Prolog's terms with tuples (immutable arrays in Python) lifts the semantics of Natlog to that of Hilog \cite{hilog}. In fact, there's an injective term algebra morphism emphasizing that semantics is still that of first-order Horn Clause logic, described in full detail in \cite{hilog}.
 
We will give the gist of it here, as an injective embedding of the basic Horn Clause subset into our tuples-based representation, which results also in a way to downgrade function symbols  to atomic constants, by lifting all parts of a compound term as marked with single functions symbol, not occurring in the term algebra (e.g.,``$\$$''). Then, we can omit the special symbol as implicit, with the immutable nested tuples representation  of our terms. As we will show later, this facilitates simple, content-driven deep indexing of ground term databases and neuro-symbolic plug-ins.

\BX Transformation to nested single function-symbol terms:\\
\verb~f(A,g(a,B),B)~$~~\Rightarrow~~$ \verb~$(f,A,$(g,a,B),B)~.
\EX
Let $T^{\$}$ be the term algebra $\mathbb{T}$ extended with the constant symbol ``$\$$''. We define a function $hl : \mathbb{T} \to \mathbb{T^{\$}}$  as follows:

\BE
\I if $c$ is a constant, then $hl(c)=c$
\I if $v$ is a variable, then $hl(v)=v$
\I if $x~=~f(x_1,\dots,x_n)$ then $hl(x)~=~ \$(f,hl(x_1),\dots,hl(x_n))$
\EE
Following \cite{padl21}, we denote $\odot$ the clause unfolding operation that, when iterated, computes the result of
LD-resolution, Prolog's specialization of SLD-resolution \cite{kowalski76,lloyd84}.
The transformation $hl:\mathbb{T} \to \mathbb{T^\$}$ is injective and it has a left inverse $hl^{-1}$.  The following relations hold:
\begin{equation}\label{unfimpl}
hl(C_1 \odot C_2) ~=~ hl(C_1) \odot hl(C_2)
\end{equation}
\begin{equation}
C_1 \odot C_2 ~=~ hl^{-1}(hl(C_1) \odot hl(C_2))
\end{equation}
where $C_1$ and $C_2$ are Horn Clauses. Observe that $\odot$ commutes with unification and consequently $hl$ and $hl^{-1}$ commute with LD-resolution.
The result follows by induction on the structure of $\mathbb{T}$ and the structure of $\mathbb{T^\$}$ for its inverse $hl^{-1}$.

As all terms have only ``$\$$`` in function symbol positions, by omitting them, one can see these terms as multi-way trees with variable or constant labels marking leaf nodes.
Abandoning explicitly marked function and predicate symbols in favor of nested tuples with constants and variables in leaf-positions preserves the semantics of Horn Clause resolution, while it can also accommodate the more general Hilog semantics, should one want to place compound terms in the first position in a nested tuple.

\section{The Proof-of-concept Python implementation}\label{poc}

We are now ready to overview the details of our proof-of-concept Natlog implementation. A simple regular expression-based 
{\em Scanner}\footnote{\url{https://github.com/ptarau/pypro/blob/master/natlog/scanner.py}}
 returns tokens matching the usual Prolog atomic terms. The recursive descent 
 {\em Parser}\footnote{\url{https://github.com/ptarau/pypro/blob/master/natlog/parser.py}} 
 handles our Horn clause syntax and the parenthesized Python-like nested tuple  notation for compound terms.

\subsection{Representing Terms with Native Python Types}
Python is a language that can be often 30-40 times slower than {\bf C} for equivalent code\footnote{
\url{https://benchmarksgame-team.pages.debian.net/benchmarksgame/fastest/python3-gcc.html}
}. At the same time it can be significantly accelerated by using the pypy JIT compiler\footnote{\url{https://www.pypy.org/}}, provided that one relies as much as possible on Python's native types that would be mapped by the JIT compiler into equivalent {\bf C} types.

As our terms are immutable and unification operations will dominate the inner-loop of the interpreter, we opt for an {\em environment} represented as a Python list (in fact, a dynamic mutable array), in which variables, represented as integer indices, will place their bindings to other variables or immutable terms. Thus, we use a simple structure-sharing representation for our terms, without having to manage our own heap. As Python's native {\tt int} type is borrowed to represent variables inside our nested tuples, we represent actual integers with a class {\tt Int} defined as a wrapper over the native {\tt int} type.

Unification\footnote{\url{https://github.com/ptarau/pypro/blob/master/natlog/unify.py}}
 (with an option for occurs-check) and trailing are implemented as usual, the unification algorithm receiving, besides the two terms, the mutable environment and the trail, as parameters.

The interpreter, kept as simple as possible in this proof-of-concept implementation, is an iteration  of the clause unfolding operation, similar to the fast Java-based system described in \cite{iclp17}, with special care for a natural embedding in the Python ecosystem.

Our data-type choices result in an order of magnitude 
speed-up\footnote{\url{https://github.com/ptarau/pypro/blob/master/small_bm.py}} with the 
pypy JIT compiler (e.g., from {\tt 35.0755} seconds to {\tt 2.9985} seconds on a 10 Queens, all solutions program).

\subsection{Quick Overview of the Natlog Interpreter} 

The Natlog interpreter\footnote{\url{https://github.com/ptarau/pypro/blob/master/natlog/natlog.py}} yields its answers directly from the stream of answers generated by the {\tt step} function which receives a list of goals to reduce. The interpreter is written as declaratively as possible, in a purely functional style, with its key steps implemented as nested inner functions.

It's code skeleton looks like this:
\begin{code}
# unfolds repeatedly; when done, yields an answer
def interp(css, goals , transformer, db=None):
  def step(goals):
    # reduces goals and yields answer when no more goals
    ...
  yield from step((goals[0], ()))
\end{code}
The work is done by the inner {\tt step} function and its inner functions, all accessing the list of clauses {\tt css}, the list of {\tt goals}, a {\tt transformer} overriding Python's {\tt eval} function for possible filtering out of insecure functions or tracing, as well as an optional ground term fact database {\tt db}.

\subsubsection{The Unfolding Step}

Goals are reduced by unifying the current goal against candidate clauses and extending, on success, the list of goals with new goals derived from the body of a matching clause.

\begin{code}
    # unfolds a goal using matching clauses
    def unfold(g, gs):
      for cs in css:
        h, bs = cs
        h=relocate(h) # create fresh head
        if not unifyWithEnv(h, g, vs, trail=trail, ocheck=False):
          undo()
          continue  # FAILURE
        else:
          bs1 = relocate(bs) # create fresh body
          bsgs = gs
          for b1 in reversed(bs1) :
            bsgs=(b1,bsgs)
          yield bsgs  # SUCCESS
\end{code}

Like in \cite{iclp17}, we  {\em relocate} the variable references in the head of the clause prototype for its tentative unification with the goal, and propagate that to its body on success, while undoing them on failure.

\subsection{The Python Calls}

\BX
The code supporting a simple call to Python (e.g., ``print''), with no return expected, looks as follows.
\begin{code}
   def python_call(g,goals):
      f=eval(g[0])
      args=to_python(g[1:])
      f(*args)
\end{code}
\EX

More general calls, receiving the value returned by a Python function are implemented with the convention that the result is unified with the last argument of the term. The same convention is followed when calling a Python generator that yields a (possibly infinite) stream of results. The {\tt step} function is called recursively (in this case in a continuation passing style), but turning it into a {\em trampoline} that eliminates stack usage is quite easy along the lines of \cite{padl21}. 
\BX
 A call that unifies  the last argument of  a term with a value yield from a generator after
 passing to it its first arguments, assumed ground looks as follows:
\begin{code}
   def gen_call(g,goals) :
      gen=transformer(g[0])
      g=g[1:]
      v=g[-1]
      args=to_python(g[:-1])
      for r in gen(*args) :
        r=from_python(r)    
        if unifyWithEnv(v, r, vs, trail=trail, ocheck=False):
          yield from step(goals) # recursive call here
        undo()
\end{code}
\EX
Note the {\tt to\_python} function that creates a fully dereferenced nested tuple representation of a term and the forwarding of yields from the {\tt step} function, in a continuation passing style.

\subsection{Key Features of the Python Embedding}\label{inter}

Calling a Python function is exposed by prefixing the function name with {\tt \#}.
\BX A simple function call.
\begin{code}
goal X : b X, #print 'printing b =' X, c X.
\end{code}
\EX

Similarly, generators can be called and have their yields collected  into a logic variable as if they were alternative bindings obtained on backtracking: 
\BX A small program exposing Python generators in Natlog code.
\begin{code}
good l.
good o.
goal X : ``iter hello X, good X.
goal X : `` range 1000 1005 X.
\end{code}
The query ``\verb~goal Answer?~'' first prints out the characters   \verb~'l','l','o'~ and then natural numbers from {\tt 1000} to {\tt 1004}.
\EX

\BX Pretending to be a Python generator.
We assume that the string ``{\tt prog}'' contains the permutation program in Example \ref{perm}.
Running the query:
\begin{code}
  n=natlog(text=prog)
  for answer in n.solve("perm (a (b (c ()))) P?"):
    print(answer[2])
\end{code}
we obtain a stream of answers yield by our interpreter pretending to be a Python generator:
\begin{codex}
('a', ('b', ('c', ())))
('b', ('a', ('c', ())))
('b', ('c', ('a', ())))
('a', ('c', ('b', ())))
('c', ('a', ('b', ())))
('c', ('b', ('a', ())))
\end{codex}
Note that in this proof-of-concept implementation list constructors are simply nested tuples of length 2 and components of the answer tuples yield by the interpreter can be accessed with the usual array index notation as in {\tt answer[2]}, selecting the argument P of the tuple \verb~(perm _  P)~.
\EX

We have also enabled {\em yielding an answer from any point in the resolution process}, as a feature enabled by coroutining with first class Prolog engines as described in \cite{bp2011} and also implemented in SWI-Prolog\footnote{\url{https://www.swi-prolog.org/pldoc/man?section=engines}}.

\BX Notation for yielding an answer from an arbitrary point in a program.
\begin{code}
n = natlog(text="worm : ^o, worm.")
for i , answer in enumerate(n.solve("worm ?")):
  print(answer[0])
  if i >= 42 : break
\end{code}
The program will yield, under Python's control of how many answers it wants from the infinite stream generated by ``{\tt worm}``, the result:
\begin{codex}
ooooooooooooooooooooooooooooooooooooooooooo
\end{codex}
\EX

This feature, in combination with Python's ability to serialize the state of our runtime system, could be used, along the lines of \cite{td:tlp} as a lightweight mobile code mechanism, given also that Natlog's code is written in pure Python, with no dependencies.

We separate the Horn Clause-equivalent rule language used for {\em code} from the {\em ground database} used to access datasets for Machine Learning applications. This allows us to import .csv, .tsv and .json files as if they were collections of facts in our ground term database.
 
\BX  
Calls to predicates in the ground database look as ordinary calls except that they are prefixed with a tilde character:
\begin{codex}
~my_database_predicate A B ... Z
\end{codex}
Natlog will handle them using specialized indexing and unification algorithms.
\EX

\section{The Content-driven Ground Database Indexer}\label{db}
 
 Traditional Prolog implementations conflate code-indexing and ground database indexing.   
 In fact, clever just-in-time indexing mechanisms (as in YAP and later SWI-Prolog) \cite{yapJIT} or trie-based indexing used for tabling in XSB-Prolog \cite{xsb} are a testimony of the implementors' ingenuity for covering both code-centered and data-centered indexing in a uniform way. Still, while traditional Prolog indexing makes sense when facts are just base cases of recursive predicates, is is likely to be suboptimal for a database of a few million possibly deep ground facts.
In particular, this is the case when small code snippets need to interact with very large datasets in Machine Learning applications, especially when the underlying logic engine, like it is the case of Natlog, is implemented as comparatively slower Python code.

Moreover, when indexing arbitrarily complex ground terms (e.g., when fetching a JSON file with a deeply nested structure),
one might want to  just focus on the {\em set of constants} occurring in a given data component. One can devise, based on them, a {\em content-driven} indexing mechanism,
and then delegate refining the correct matches to a specialized unification algorithm.

Natlog's  content-driven indexing mechanism\footnote{\url{https://github.com/ptarau/pypro/blob/master/natlog/db.py}} is specialized to ground terms. It is kept separate from the logic engine and it can be  overridden by more elaborate indexing mechanisms, including neural-networks in inference-mode.

\subsection{The Indexing Mechanism} 
 
When adding a fact to the ground database represented as  nested tuple, with atomic constants occurring as leaves, we index it for the set of constants occurring in it. We use for that a Python dictionary  that associates to each constant the set of clauses in which the constant occurs.

Given a query (possibly containing variables), we  compute all its ground matches with the database, knowing that {\em if a constant occurs in the query, it must also occur in a ground term that unifies with it}, as the ground term has no variables in any position that would match the constant otherwise.

We start with the set of clauses where the first constant occurs. Then we reduce it progressively by intersecting it with the sets of clauses in which subsequent constants in the query occur, as shown in the following Python code snippet:
\begin{code}
  def ground_match_of(self,query):
    # find all constants in query
    constants=const_of(query)
    if not constants :
      # match against all clauses self.css, no help from indexing
      return set(range(len(self.css)))
    # pick a copy of the first set where c occurs
    first_constant=next(iter(constants))
    matches=self.index[first_constant].copy()
    # shrink it by intersecting with sets  where other constants occur
    for x in constants:
      matches &= self.index[x]
    # these are all possible ground matches - return them
    return matches
\end{code}
Simple optimizations  include selecting the constant with the fewest occurrences to provide the set to start with.    
 
\subsection{Specializing Unification against Ground Terms}\label{gunif}

The indexing mechanism relies on the following facts about unifying against a ground term:
\BI
\I unification against a ground term is sound without occurs-check, given that variables occur only on one side
\I if a constant occurs in the (possibly non-ground) query, it must also occur in
    a ground term that unifies with it, as the ground term
    has no variables that would match the constant otherwise
\EI
The indexing mechanism returns a set of ground fact candidates but does not guaranty unifiability, which depends not only on the constant set occurring in ground terms but also on their tree structure. Consequently, we rely on the usual unification step of LD-resolution to refine the set of terms that passed the indexing test and filter out false positives.

 \subsection{Extensions}
 
 \subsubsection{The Path-to-a-constant Indexing Mechanism}
 To approximate more closely unification against a ground term, the indexing algorithm will use the path to the location of each constant, represented as an immutable tuple (to ensure that it is ``hashable'' -- a requirement for keys in Python's dictionaries).
 \BX Paths to constants represented as tuples in a nested tuple representing a ground term.
 
 \noindent term: $T$ = \verb~(f a (g (f b) c))~\\
 paths: \verb~(0 f) (1 a) (2 0 g) (2 1 0 f) (2 1 1 b) (2 2 c)~
 
 \EX
 It is easy to see that all the properties of unification against a ground term mentioned in subsection \ref{gunif} extend to the case of a path-to-constant indexing mechanism.
 
 Thus, for  large fact databases containing deep ground terms, one  can refine indexing by using as a key the exact path locating the constant (a leaf in the multi-way tree representation of a term).
 
 \subsubsection{Structure Matching}
 Starting with a term like  $T$ = \verb~(f a (g (f b) c))~, if one replaces each constant with ``\verb~o~``, the structure of the term can be succinctly described as \verb~(oo(o(oo)o))~ and compactly stored. A query term  containing variables like $Q$=\verb~(f a (g X c))~ could then be described as (oo(o*o)) with ``\verb~*~'' marking variables. Then, as a pre-unification step, candidates can be scanned knowing that '*' must match the corresponding fully parenthesized expression, given the fact that the  tree associated to $Q$ must be a subtree of the tree associated to  $T$.

 \subsubsection{The Ground term Database as an Associative Memory}
 
Another extension is the possible case of ground clauses as  ``associative memory'' elements. This fits LD-resolution semantics as
content-driven ground indexing of the heads would yield, besides bindings, clause bodies to be further explored via the usual LD-resolution steps.

\section{Using a Neural Network Plug-in as a Content-Driven Ground Term Database Indexer}\label{nn}

A neural-net based equivalent of our content-driven indexing algorithm is obtained by overriding its database constructor
with a neural-net trained database {\tt ndb()} as shown below:
\begin{code}
class neural_natlog(natlog):
  def db_init(self):
    self.db=ndb() # neural database equivalent
\end{code}
Otherwise, the interface remains unchanged, the LD-resolution engine being oblivious to working with the ``symbolic'' or ``neural'' ground-fact database.

We keep the dependencies of Natlog to only two Python packages, {\tt numpy} and {\tt scikit-learn}, and these dependencies are only activated for {\tt neural\_natlog}.

The code skeleton for the neural ground term database\footnote{\url{https://github.com/ptarau/pypro/blob/master/natlog/ndb.py}}
 is implemented as the {\tt ndb} class below:
\begin{code}
class ndb(db) :
  def load(self,fname,learner=neural_learner):
    # overrides database loading mechanism to fit learner
    ...
    
  def ground_match_of(self,query_tuple):
    # overrides database matching with learned predictions
   ...
\end{code}
The overridden {\tt load(...)} method will fit a {\tt scikit-learn}  machine learning algorithm\footnote{abbreviated {\tt sklearn} when imported as a package} (by default a simple multi-layer perceptron neural network), to yield, when used in inference mode by the method {\tt ground\_match\_of(...)}, the set of ground clauses likely to  match the query. 

As in the case of the  content-driven ground term indexer, we will create an association between the set of constants occurring in the query and the set of ground facts containing them in the database.

Our design will keep in mind the need to return more than one answer from the neural indexer, as multiple facts can match a given query. As usual in Machine Learning terminology, we  denote {\bf X} the input from which the algorithm will need to learn the expected output, denoted {\bf y}. Our model will be a {\em classifier} that associates to each constant the set of clauses it occurs in, to mimic the ground fact database indexer that it overrides. 

The training mode, happening in the {\tt load(...)} method,  proceeds as follows:
\BE
\I load the dataset from a Natlog, .csv, .json file
\I have the superclass ``{\tt db}'' create the index associating to each constant the set of facts it occurs in
\I create a {\tt numpy} diagonal matrix with one row for each constant (our {\bf X} array)
\I compute a {\em OneHot encoding}\footnote{basically a bitvector of fixed size} for the set of clauses associated to each constant (our {\bf y} array)
\I call the {\tt fit} method of the the sklearn classifier (a neural net by default, but swappable to any other, e.g., Random Forest, Stochastic Gradient Descent, etc.) with  the {\bf X,y} training set
\EE

The inference mode, happening in the {\tt ground\_match\_of(...)} method proceeds as follows:
\BE
\I compute the set of all constants in the query that occur in the database
\I compute their OneHot encoding
\I use the classifier's {\tt predict} method to return a bitset encoding the predicted matches
\I decode the bitset to integer indices in the database and return them as matches
\EE

\noindent The following  examples illustrate some typical use cases.

\BX Natlog program calling a database of properties of chemical elements (note the \verb|~| prefix in the first clause).
\begin{code}
data Num Sym Neut Prot Elec Period Group Phase Type Isos Shells : 
   ~ Num Sym Neut Prot Elec Period Group Phase Type Isos Shells.

an_el Num El	: data Num El '45' '35' '35' '4' '17' liq 'Halogen' '19' '4'.
gases Num El : data Num El  _1   _2   _3  _4   _5 gas  _6        _7   _8.
\end{code}
\EX
\BX
The ground database loaded from the tab-separated (.tsv) file {\tt elements.tsv}:
\begin{codex}
1	H	0	1	1	1	1	gas	Nonmetal	3	1
2	He	2	2	2	1	18	gas	Noble Gas	5	1
3	Li	4	3	3	2	1	solid	Alkali Metal	5	2
...
84	Po	126	84	84	6	16	solid	Metalloid	34	6
85	At	125	85	85	6	17	solid	Noble Gas	21	6
86	Rn	136	86	86	6	18	gas	Alkali Metal	20	6
\end{codex}
\EX

\BX
The Python program\footnote{\url{https://github.com/ptarau/pypro/blob/master/tests.py}} running the Natlog code and the neural-net classifier:
\EX
\begin{code}
def ndb_chem() :
  nd = neural_natlog(
    file_name="natprogs/elements.nat",
    db_name="natprogs/elements.tsv"
  )
  print('RULES');print(nd)
  print('DB FACTS');print(nd.db)
  nd.query("an_el Num Element ?")
  nd.query("gases Num Element ?")
\end{code}
will print out (after listing its data and code), as the result of the second query, the atoms that occur as gases at normal temperature ranges, all computed as candidates provided by the {\em neural} indexer and then validated by a {\em symbolic} unification step:

\begin{codex}
GOAL PARSED: (('gases', 0, 1),)
ANSWER: ('gases', '1', 'H')
ANSWER: ('gases', '2', 'He')
...
ANSWER: ('gases', '54', 'Xe')
ANSWER: ('gases', '86', 'Rn')
\end{codex}

\section{Related Work}\label{rel}

A similar syntactic departure from the function symbol followed by its parenthesized arguments was present  in functional programming as far as in LISP and has persisted in Miranda and Haskell. It has also propagated to Micro-Prolog \cite{microprolog}.
 The syntax and Hilog-like semantics of Natlog is similar to that of the Java-based iProlog\footnote{\url{https://github.com/ptarau/iProlog}} system
\cite{iclp17}.  Derived from "first principles" (starting with a Prolog meta-interpreter), iProlog implements LD-resolution with help of a goal-stack, itself similar to the effect of the binarization transformation described in \cite{bp2011}. Contrary to the emphasis  on performance in the case of iProlog, which despite being an interpreter and implemented in Java, is only 2-3 times slower than C-based compiled Prolog implementations, our focus in Natlog was its seamless integration with the Python machine-learning ecosystem. This task was largely facilitated by the presence of Python's {\tt eval} function, providing easy bi-directional calls. To keep things simple, Natlog adopts a structure-sharing execution algorithm, while iProlog is essentially, like the WAM \cite{kaci91:WAM}, a structure-copying system.

Work on deep indexing of Prolog terms  \cite{ramesh} used an automaton based on paths tracing function symbols in the term tree. That is similar to  the path-to-the-constant indexing extension discussed in this paper, except for the simplifications that we obtain by focusing on ground terms and the expression of our algorithm exclusively in terms of set intersection operations.   
In \cite{iclp17} a more elaborate indexing mechanism  relying on constant symbols occurring in a term is described, which also accommodates non-ground clauses. The mechanism adopted in Natlog can be seen as its specialization restricted to ground terms. 

More radical departures from the traditional function-symbol + arguments representation of Prolog terms are explored in \cite{padl21} via a series of unification-oblivious program transformations, that generalize the assumptions on which Hilog \cite{hilog} and (in a syntactically different wrapping)  Natlog rely.

It is not unusual for logic programming languages to adopt a tight integration with Python as is the case with the ASP system
clingo\footnote{\url{https://potassco.org/clingo/}} \cite{clingo} or with Problog\footnote{\url{https://dtai.cs.kuleuven.be/problog/}} \cite{deepProblog}, where neural machine learning is  integrated  with a probabilistic logic programming system.

We refer to \cite{d2020neurosymbolic} for an overview of of the fast-growing attempts to neuro-symbolic integration and to 
\cite{lamb2020graph} for an overview specialized to graph-neural networks, especially suitable for relational inference on unstructured data.
In \cite{de2020statistical} a large number of neuro-symbolic systems are surveyed, along multiple features, among which we mention the focus on the expressiveness of the logic (ranging from propositional to first order predicate logic) and its probabilistic or deterministic nature.
An ontology of design patterns for neuro-symbolic systems is explored in \cite{van2021modular}, to which our neural ground fact database indexer would be a possibly new addition.

\section{Conclusions}\label{conc}
We have described informally (but as informatively as possible) the Natlog logic programming language and its proof-of-concept Python implementation. We will highlight here the novel ideas that it brings, with focus on its possible practical contributions to embedding logic programming tools and techniques in deep-learning applications.

The tight integration with Python's generators and  coroutining mechanisms enables extending machine-learning applications with an easy to grasp logic programming subsystem.
Our departure from traditional Prolog's predicate and term notation puts forward a more readable syntax together with a more flexible Hilog-like semantics. Its closeness to natural-language sentences is likely to be an incentive for adoption by data-scientists not familiar with logic programming.

The content-driven indexing against ground term fact databases is new and it is a potentially useful addition to Prolog and Datalog systems, especially in its extended path-to-the-constant form. As we have shown in a Python code snippet, it is also easily implementable.

Our neural-net plugin mechanism offers a simple yet practical way to integrate deep-learning and logic-based inferences. 
It is also a new way to approach neuro-symbolic programming by delegating to the machine learning ecosystem a simple subtask at which it is good, while validating correctness of its results symbolically as part of Prolog's well-known LD-resolution execution mechanism.

\bibliographystyle{eptcs}
\bibliography{theory,tarau,proglang,biblio}

\end{document}